\documentclass[pre,twocolumn,superscriptaddress,showpacs,floatfix]{revtex4}
\usepackage{color,epsfig}
\usepackage{bm}
\usepackage{amsmath,amsfonts,amssymb}
\DeclareMathOperator{\sgn}{sgn}
\usepackage{natbib}

\begin{document}

\title{Magneto-strain-driven quantum engine on a graphene flake}
\author{Francisco J. Pe\~{n}a}
\affiliation{Instituto de F\'isica, Pontificia Universidad Cat\'olica de Valpara\'iso, Av. Brasil 2950, Valpara\'iso, Chile.}
\author{Enrique Mu\~{n}oz}
\email{munozt@fis.puc.cl}
\affiliation{Facultad de F\'isica, Pontificia Universidad Cat\'olica de Chile, Vicu\~{n}a Mackenna 4860, Santiago, Chile.}
\affiliation{Research Center for Nanotechnology and Advanced Materials CIEN-UC, Pontificia Universidad Cat\'olica de Chile, Santiago, Chile.}

\date{\today}

\begin{abstract}
We propose a novel conceptual design for a graphene-based quantum engine, driven by a superposition of mechanical strain and an external magnetic field. Engineering of strain in a nanoscale graphene flake creates a gauge field with an associated uniform pseudo-magnetic field. The strain-induced pseudo-magnetic field can be combined with a real magnetic field, leading to the emergence of discrete relativistic Landau levels within the single-particle picture. The inter-level distance and hence their statistical population can be modulated by quasi-statically tuning the magnetic field along a sequence of reversible transformations that constitute a quantum mechanical analogue of the classical Otto cycle.
\end{abstract}


\maketitle

\section{Introduction}
The concept of a quantum heat engine (QHEN) has been discussed extensively in the literature \cite{Bender_02,Bender_Brody_00,Wang_011,Wang_He_012,Quan_06,Arnaud_02,Latifah_011,Quan_Liu_07,
Scully_03,Scully_011,Quan_Zhang_05,Dong_2013,Dong_PRB_2013,Hubner_2014} as an
alternative to efficiently recover, on a nanoscale device, thermal energy in the form of useful work. In a QHEN, in contrast with a classical heat engine, the working substance exhibits quantum mechanical properties. Interesting examples of this concept are constituted by photosynthesis in plants \cite{Dorfman_013}, as well as human-designed photocells \cite{Scully_03,Scully_011}, where the working substance are thermalized photons. Moreover, it has been recently proposed that if the reservoirs are also of quantum mechanical nature, these could be prepared
into quantum coherent states \cite{Scully_03,Scully_011} or into squeezed thermal states \cite{Lutz_014}, thus allowing for an enhancement of
the theoretical photocell efficiency beyond the classical Carnot limit \cite{Lutz_014,Scully_011,Scully_03}. These examples by no means constitute the only possible configurations, since a number of different designs based on alternative principles have been proposed in the literature, such as entangled states in a qubit \cite{Huang_013} and quantum mechanical versions of the Diesel \cite{Dong_PRB_2013} and the Otto cycle \cite{Li_Negentropy_013,Lutz_014,Hubner_2014}. At the conceptual level, a statistical ensemble of confined single-particle systems can undergo a cycle of reversible transformations driven by a generalized external field. The driving field can be a mechanical force \cite{Wang_He_012,Wang_011,Bender_02,Bender_Brody_00,Latifah_011} that modifies the inter-level spacing of the single-particle spectrum, thus inducing a sequence of transitions on the statistical population of the single-particle states, in close analogy to a classical gas confined by a piston. We have discussed generalizations of this idea in the context of relativistic Dirac particles \cite{Munoz_Pena_012}. More recently, we proposed \cite{Munoz_Pena_014} a magnetically driven 
QHEN, based on the combined effects of a parabolic confining potential, representing a semiconductor quantum dot, and an external magnetic field. In the single-particle picture, this configuration possesses an exact solution in terms of effective Landau levels, which constitute a discrete spectrum, where the inter-level distance can be modulated by tuning the external magnetic field \cite{Munoz_Pena_014}. 

In the present work, we propose yet a different alternative, through the conceptual design of a QHEN based on a strained single-layer graphene flake.
It has been widely discussed in the literature \cite{Guinea_Nat_2010,Guinea_PRB_2010,Levy_2010, deJuan_2013} that within the single-particle picture provided by the effective Dirac
theory, mechanical strain in graphene translates into a gauge field whose curl plays the role
of a pseudo-magnetic field. Beyond theoretical calculations, this effect has been experimentally observed \cite{Levy_2010} through the evidence of pseudo-relativistic Landau levels in strained graphene, in the absence of an actual magnetic field. Moreover, via "strain-engineering" \cite{Guinea_Nat_2010,Guinea_PRB_2010} it is possible to create a mechanical strain field that induces a nearly uniform pseudo-magnetic field up to $B_{S}\sim 300$ T \cite{Levy_2010}. On top of this strain field, it is of course possible to impose a real uniform magnetic field perpendicular to the plane of the sample, with the expected additive effect. In this work, we propose a sequence of reversible transformations, involving the quasi-static tuning of the external magnetic field, that constitute a quantum mechanical analogue of the classical Otto cycle. We calculate the efficiency of this cycle, and compare with the Carnot efficiency at the same temperatures of the source and sink thermostats. In particular, we show that better performance is obtained by combining magnetic and strain fields than by adjusting strain solely.

\begin{figure}[tbp]
\centering
\epsfig{file=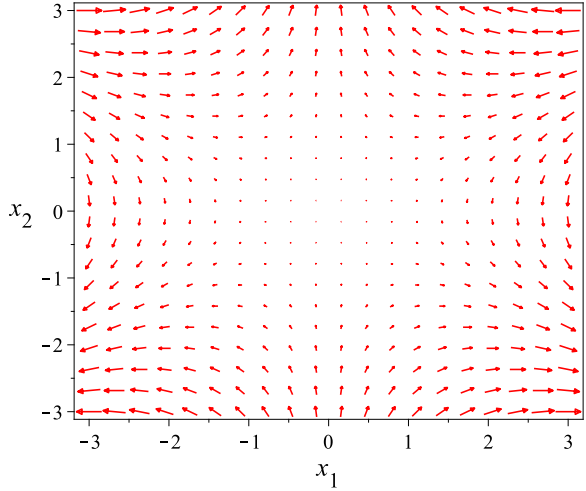,width=0.9\columnwidth,clip=}
\caption{(Color online) The deformation vector field $\mathbf{u}(x_{1},x_{2}) = (u_{1},u_{2},0)$ defined by Eq.(\ref{eq6}).}
\label{fig1}
\end{figure}
\section{Strained graphene in an external magnetic field}

The effect of strain-induced pseudo-magnetic fields on the electronic properties of graphene has been extensively discussed in the  literature \cite{Castro_Neto09,Guinea_Nat_2010,Guinea_PRB_2010,Levy_2010, deJuan_2013}. Moreover, by strain engineering it is possible to generate nearly homogeneous pseudo-magnetic fields \cite{Guinea_Nat_2010,Guinea_PRB_2010}, thus allowing for the emergence of relativistic Landau levels, as confirmed experimentally \cite{Levy_2010}. The components of the in-plane pseudo-vector potential $\mathbf{A}_{S}$ induced by mechanical strain are \cite{Guinea_Nat_2010,Guinea_PRB_2010,deJuan_2013}
\begin{eqnarray}
A_{S,1} = \frac{\beta}{2a}\left( u_{11} - u_{22} \right),\,\,\,\,A_{S,2} = \frac{\beta}{2a}\left( -2 u_{12}\right).
\label{eq1}
\end{eqnarray}
Here, the local displacement vector is defined as $\mathbf{u} = (u_{1},u_{2},z)$, with $u_{i}$ the in-plane
components, and $z$ the out-of-plane component, while $\beta = \partial\ln t/\partial\ln a$ is the relative change of the hopping parameter $\delta t/t$ with respect to the dilation of the lattice constant $\delta a/a$ \cite{Goerbig_2011}. The strain tensor $[u_{ij}]$ is defined by 
\begin{eqnarray}
u_{ij} = \frac{1}{2}\left( \partial_{i}u_{j} + \partial_{j} u_{i} + \partial_{i}z\partial_{j}z  \right).
\label{eq2}
\end{eqnarray}
It was recently shown \cite{deJuan_2012,deJuan_2013,Manes_2013} that the continuum expansion of the tight-binding Hamiltonian for strained graphene in the vicinity of a single Dirac point (valley), up to $O(u_{ij}^{2})$, is given by
the expression
\begin{eqnarray}
\hat{H} &=&  -i\int d^{2}x\hat{\psi}^{\dagger}(x)\left[ v_{ij}(x)\hat{\sigma}_{i}\partial_{j}
+ i v_{F}\hat{\sigma}_{i}A_{S,i}\right.\nonumber\\
&&\left. + v_{F}\hat{\sigma}_{i}\Gamma_{i}  -\bar{\gamma} B_{S} \hat{\sigma}_{3}  \right]\hat{\psi}(x).
\label{eq3}
\end{eqnarray}
Here, $v_{F}\sim 10^{6}\,m\,s^{-1}$ is the Fermi velocity for the undistorted graphene lattice. The presence of strain induces a local position dependence on the velocity \cite{deJuan_2012,deJuan_2013}, which then becomes a tensor 
\begin{eqnarray}
v_{ij}(x) = v_{F}\left( \delta_{ij} -\frac{\beta}{4}\left( 2 u_{ij} + \delta_{ij} u_{kk} \right) + \tilde{u}_{ij} \right).
\label{eq4}
\end{eqnarray}
De Juan {\it{et al.}} \cite{deJuan_2013} have shown that in order to describe correctly the effects of strain, the Hamiltonian
should be expressed in the so-called "laboratory" frame, that is with all the fields described with respect to the instantaneous atomic positions in the deformed crystal configuration $y_{i} = x_{i} + u_{i}(x)$. This statement leads to the inclusion of the last contribution into Eq.(\ref{eq4}) for the velocity tensor \cite{deJuan_2013}, given by $\tilde{u}_{ij} = \frac{1}{2}\left( \partial_{i}u_{j} + \partial_{j}u_{i} \right)$. On top of the gauge field $\mathbf{A}_{S}$, a "geometric" vector field $\Gamma_{i}$ appears, whose origin can be understood within the covariant picture \cite{deJuan_2012} as the pseudo-spin connection for fermions propagating in a curved space. When expressed in the "laboratory" frame \cite{deJuan_2013},
it becomes
\begin{eqnarray}
\Gamma_{i} = \frac{1}{2v_{F}}\partial_{j}v_{ij} = -\frac{\beta}{4}\left( \partial_{j}u_{ij} + \frac{1}{2}\partial_{i}u_{jj}  \right) + \frac{1}{2}\partial_{j}\tilde{u}_{ij}.
\label{eq5}
\end{eqnarray}

The pseudo-magnetic field generated by the strain-induced gauge potential is given by
$\mathbf{B}_{S} = \hat{e}_{3}\left( \partial_{1}A_{S,2} - \partial_{2}A_{S,1} \right)$. The last term in the
Hamiltonian Eq.(\ref{eq3}) represents a pseudo-Zeeman coupling between the pseudo-spin
degree of freedom and the strain pseudo-magnetic field $B_{S}$ \cite{Manes_2013}.

Let us consider the deformation field (see Fig.\ref{fig1})
\begin{eqnarray}
u_{1} = -2 u_{S} x_{1}x_{2},\,\,\,u_{2} = -u_{S}(x_{1}^{2} - x_{2}^{2}),\,\,\,z=0,
\label{eq6}
\end{eqnarray}
where $u_{S}$
is a constant characterizing the magnitude of the in-plane displacement, in association through Eq.(\ref{eq1}) with the vector potential
\begin{eqnarray}
\mathbf{A}_{S} = 2\frac{\beta u_{S}}{a}\left( -x_{2}, x_{1}, 0\right).
\label{eq7}
\end{eqnarray}
This in turn implies a pseudo-magnetic field
\begin{eqnarray}
\mathbf{B}_{S} = 4\frac{\beta u_{S}}{a}\hat{e}_{3},
\label{eq8}
\end{eqnarray}
which is constant in magnitude and points along the out-of-plane direction $\hat{e}_{3}$. 
Let us now discuss what is the effect of this particular deformation on the local Fermi velocity. First, it is straightforward to check that $u_{kk} = \nabla\cdot\mathbf{u} = 0$,
and that since $z=0$, we have $\tilde{u}_{ij} = u_{ij}$. 
The local Fermi velocity tensor
is then directly calculated by substituting these relations into Eq.(\ref{eq4}) 
\begin{eqnarray}
v_{ij} = v_{F}\left(\delta_{ij} -\frac{\beta}{2} u_{ij} + u_{ij} \right) = v_{F}\delta_{ij},
\label{eq9}
\end{eqnarray}
where in the last step we used that for graphene $\beta = \partial\ln t/\partial\ln a \sim 2$ \cite{deJuan_2013, Goerbig_2011}. Therefore, for the deformation field defined by Eq.(\ref{eq6}), the velocity is a constant diagonal tensor
$v_{ij} = v_{F}\delta_{ij}$.

Finally, let us analyze the corresponding expression for the field $\Gamma_{i}$ defined by Eq.(\ref{eq5}). After Eq.(\ref{eq6}) we have $u_{jj} = 0$, and hence $\partial_{i}u_{jj} = 0$. Moreover, 
it is straightforward to verify from Eq.(\ref{eq6}) that $\partial_{j}u_{ij} = \partial_{j}\tilde{u}_{ij} = 0$. Therefore, for the deformation field defined by Eq.(\ref{eq6}) we have $\Gamma_{i} = 0$, and the Dirac Hamiltonian Eq.(\ref{eq3}) reduces to
\begin{eqnarray}
\hat{H} =  \int d^{2}x\, \hat{\psi}^{\dagger}(x) \left\{v_{F} \hat{\boldsymbol{\sigma}}\cdot\left[-i\nabla + \mathbf{A}_{S} \right] -  \bar{\gamma} B_{S} \hat{\sigma}_{3} \right\}\hat{\psi}(x),\nonumber\\ 
\label{eq10}
\end{eqnarray}
with $\hat{\boldsymbol{\sigma}} = \left(\hat{\sigma}_{1}, \hat{\sigma}_{2} \right)$ and $\mathbf{A}_{S}$ given by Eq.(\ref{eq7}).

The Hamiltonian in Eq.(\ref{eq10}) describes the physics in the vicinity of a single Dirac point (valley). The inclusion of the
pseudo-magnetic field terms breaks the valley degeneracy since, as opposed to a real magnetic field, the strain-induced gauge potential has opposite signs in the vicinity of each Dirac point \cite{Castro_Neto09}. Therefore, we define the Hamiltonian describing both valleys by the higher-dimensional structure
\begin{eqnarray}
\hat{H} = \int d^{2} x \hat{\Psi}^{\dagger}(x)\left[\begin{array}{cc}\hat{H}^{+} & 0\\0 &\hat{H}^{-}  \end{array} \right]\hat{\Psi}(x).
\label{eq11}
\end{eqnarray}
Here, $\xi=\pm$ represents each of the $K_{\xi} $ valleys, corresponding to the two inequivalent points $\xi \frac{4\pi}{3\sqrt{3}a}\hat{\mathbf{e}}_{1}$ in the first Brillouin zone, respectively.
By adopting the standard convention for the components of the spinor field at each sublattice $(A,B)$
\begin{eqnarray}
\hat{\Psi}(x) = \left(\psi_{A}^{(+)}, \psi_{B}^{(+)},\psi_{B}^{(-)},\psi_{A}^{(-)}\right)^{T} = \left(\hat{\psi}^{(+)},\hat{\psi}^{(-)} \right)^{T}
\label{eq12}
\end{eqnarray}
the effective Hamiltonian at each valley, according to Eq.(\ref{eq11}) is given by
\begin{eqnarray}
\hat{H}^{\xi} &=&  \int d^{2}x\, \hat{\psi}^{(\xi)\dagger}(x) \left\{\xi v_{F} \hat{\boldsymbol{\sigma}}\cdot\left[-i\nabla + \xi\mathbf{A}_{S} \right]\right.\nonumber\\
&&\left. -  \xi \bar{\gamma} B_{S} \hat{\sigma}_{3} \right\}\hat{\psi}^{(\xi)}(x).
\label{eq13}
\end{eqnarray}

Let us now consider the combined effect of a uniform magnetic field $\mathbf{B} = \hat{\mathbf{e}}_{3}B$
in addition to the strain-induced pseudo-magnetic field defined defined by Eq.(\ref{eq8}). Choosing
for convenience the gauge
\begin{eqnarray}
\mathbf{A} = \frac{B}{2}\left(-x_{2}, x_{1}, 0 \right),
\label{eq14}
\end{eqnarray}
we have that the effective Hamiltonian at each valley $K_{\xi}$, including the electronic spin degree of freedom becomes
\begin{eqnarray}
\hat{H}^{\xi} &=& \int d^{2}x \hat{\Psi}^{(\xi)\dagger}(x)\left\{\xi v_{F} \left(\mathbf{1}\otimes\hat{\boldsymbol{\sigma}}\right)\cdot\left[-i\nabla + \xi \mathbf{A}_{S} + \mathbf{A}\right]\right.\nonumber\\
&&\left.-  \xi \bar{\gamma} B_{S} \left(\mathbf{1}\otimes\hat{\sigma}_{3} \right)- \gamma B \left(\hat{\sigma}_{3}\otimes\mathbf{1}\right)  \right\}\hat{\Psi}^{(\xi)}(x).
\label{eq15}
\end{eqnarray}
Here, we have defined the spinor
\begin{eqnarray}
\hat{\Psi}^{(\xi)}(x) = \left(\psi_{A\uparrow}^{(\xi)},\psi_{B\uparrow}^{(\xi)},\psi_{A\downarrow}^{(\xi)},\psi_{B\downarrow}^{(\xi)} 
\right)^{T } = \sum_{s=\uparrow,\downarrow}\chi_{s}\otimes\hat{\Psi}_{s}^{(\xi)}
\label{eq16}
\end{eqnarray}
where $\hat{\psi}_{s}^{(\xi)}$ is the two-component pseudo-spinor at valley $K_{\xi}$ arising from the bipartite graphene lattice, whereas  the electronic spinors $\chi_{\uparrow} = (1,0)^{T}$ and $\chi_{\downarrow} = (0,1)^{T}$ are eigenvectors of $\hat{\sigma}_{3}$ with eigenvalues $s=\{\pm\}$. The last term in the Hamiltonian Eq.(\ref{eq15}) is the Zeeman
interaction, characterized by a coupling constant $\gamma = g\mu_{B}/2$, with $g\sim 1.8$ for graphene \cite{Zhang_2006}. Notice that the Zeeman coupling only involves the real
magnetic field $\mathbf{B}$, since the strain-induced pseudo-magnetic field $\mathbf{B}_{S}$ does not interact
with the electronic spin.

The Hamiltonian Eq.(\ref{eq15}) determines a system of two decoupled Dirac equations, one for each electronic spin component $s=\{\pm\}$,
\begin{eqnarray}
&&\left(\xi v_{F}\hat{\boldsymbol{\sigma}}\cdot\left[-i\nabla + \xi \mathbf{A}_{S} + \mathbf{A} \right]- s\gamma B \mathbf{1}-
\xi \bar{\gamma}B_{S}\hat{\sigma}_{3}\right)\psi_{s}^{(\xi)}(x)\nonumber\\
&& = E_{s}^{\xi}\psi_{s}^{(\xi)}(x).
\label{eq17}
\end{eqnarray}

The pseudo-spinor eigenstates of the Hamiltonian in Eq.(\ref{eq17}) are $\psi_{\lambda,n,m,s}^{(\xi)}(x) \equiv \langle x | \psi_{\lambda,n,m,s}^{(\xi)} \rangle$, with (see Appendix A for details)
\begin{eqnarray}
|\psi_{\lambda,n,m,s}^{(\xi)} \rangle &=& \frac{1}{\sqrt{2}}\left(\begin{array}{c}  \alpha_{n}| n-1,m \rangle \\   \beta_{n}|n,m\rangle \end{array}\right),\,\,\,\, n > 0,\nonumber\\
|\psi_{n=0,m,s}^{(\xi)} \rangle &=& \left(\begin{array}{c}  0 \\   |0,m\rangle \end{array}\right),\,\,\,\, n = 0.
\label{eq18}
\end{eqnarray}
Here, the coefficients $\alpha_{n}$ and $\beta_{n}$ are as defined in Appendix A (see Eq.(\ref{eqA13b})).
The corresponding energy eigenvalues are 
\begin{eqnarray}
E_{n,s}^{\xi} = \left\{ \begin{array}{cc}\lambda\hbar\Omega_{\xi}\sqrt{n + \Delta_{\xi}^{2}} - s\gamma B, & n > 0,\\
\xi \hbar\Omega_{\xi} \Delta_{\xi} - s\gamma B, & n=0.\end{array}\right.
\label{eq19}
\end{eqnarray}
Here, $\lambda = \pm$ is the band index, while $\Omega_{\xi}= v_{F}\sqrt{2 e |B_{\xi}|/\hbar}$ is the effective frequency, expressed in terms of the 
effective "total" magnetic field $B_{\xi} = B + \xi B_{S}$ that
results from the combination of the strain-induced pseudo-magnetic field $B_{S}$ and the real magnetic field $B$ at each valley $K_{\xi}$. We have also defined the dimensionless parameters 
$\Delta_{\xi} = \bar{\gamma} B_{S} \hbar^{-1}\Omega_{\xi}^{-1}$ at each valley. The two quantum numbers $(n,m)$ correspond to the quantization of the orbital Landau level $n\ge 0$, and the {\it{guiding-center}} (i.e. the center of the classical cyclotronic orbit) coordinate $m\ge 0$, respectively \cite{Goerbig_2011}. The energy levels
described by Eq.(\ref{eq19}) are degenerate in the guiding-center quantum number $m$, with the same degeneracy factor $N_{n,s}^{\xi} =  N_{\phi}^{\xi}$ for each Landau level and each valley. On the other hand, $N_{\phi}^{\xi} = \Phi_{B_{\xi}}/\phi_{0}$ is the number of magnetic flux quanta $\phi_{0} = h/2e$ piercing the area $\mathcal{A}$ of the graphene flake \cite{Goerbig_2011}, with $\Phi_{B_{\xi}} = B_{\xi} \mathcal{A}$ the "total" flux. As previously discussed, only the magnetic field $B$ couples to the electronic spin $s = \{\pm\}$, as seen in the Zeeman term in Eq.(\ref{eq19}), which is proportional to $\gamma=g\mu_{B}/2$. Regarding the pseudo Zeeman term, when $B_{S}$ is expressed in Tesla, we estimate \cite{Levy_2010,Guinea_Nat_2010,Manes_2013} $\bar{\gamma} = \frac{3\pi a^{3}}{4\beta}\frac{V'}{\phi_{0}} = 9.788\times 10^{-5}\,{\rm{eV\, T^{-1}}}$, for $a=1.42\, {\rm{\AA}}$ the carbon-carbon bond length, and $V' = 6\, {\rm{eV \AA}}^{-1}$ ~\cite{Manes_2013,Ferone_2011}. Remarkably, $\bar{\gamma} \sim 1.7 \mu_{B}$ is on the order of magnitude of the Bohr magneton.

\begin{figure}[tbp]
\centering
\epsfig{file=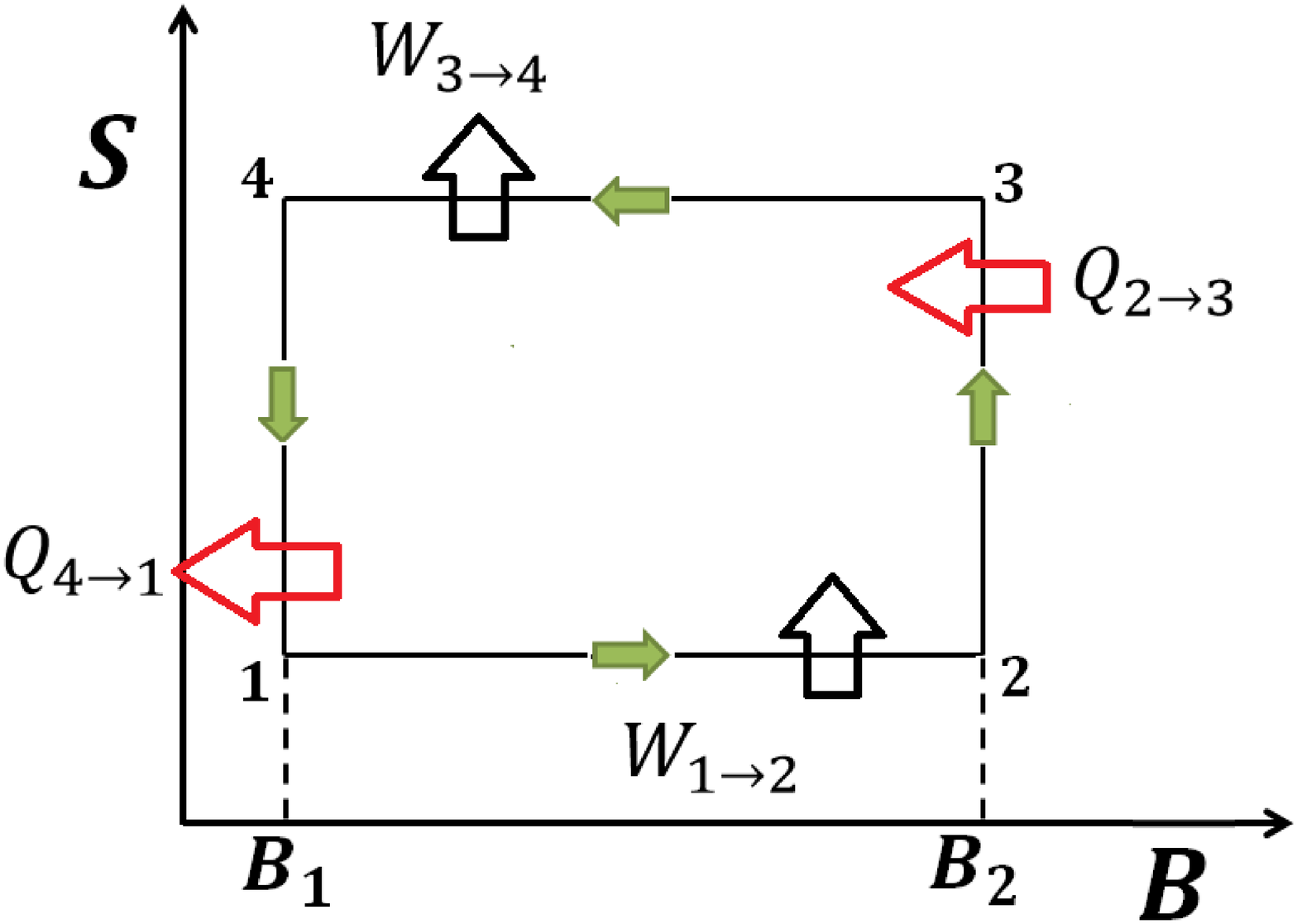,width=0.9\columnwidth,clip=}
\caption{(Color online) The cycle is pictorially represented in the entropy (S) versus external magnetic field (B) coordinates. The cycle is
composed by two iso-entropic trajectories, and two trajectories at constant external magnetic field. The {\it{cold}} reservoir is at $T_{1} = T_{C}$, whereas the {\it{hot}} reservoir is at $T_{3} = T_{H}$.}
\label{fig2}
\end{figure}

\section{Magnetic quantum engine}

As the "working substance" for our quantum engine, let us consider a statistical ensemble
of replicas of a single-particle system \cite{Munoz_Pena_012,Munoz_Pena_014} consisting
on a single electron in the conduction band ($\lambda = +$) of to the graphene flake described by Eq.(\ref{eq15}). This can in principle be achieved by charging an otherwise neutral graphene flake with a positive gate potential.
Each replica \cite{Munoz_Pena_012,Munoz_Pena_014} may be in any of the different eigenstates
of the Hamiltonian Eq.(\ref{eq17}).
The single-particle system
is then in a statistically mixed quantum state \cite{vonNeumann}, described by the density
matrix operator $\hat{\rho} = \sum_{n,m,s,\xi}p_{n,m,s,\xi}(B)|\psi_{n,m,s}^{(\xi)}(B)\rangle\langle\psi_{n,m,s}^{(\xi)}(B)|$, with $|\psi_{n,m,s}^{(\xi)}(B)\rangle$
a spinor eigenstate of the single-particle Hamiltonian Eq.(\ref{eq17}) for a given magnetic field intensity $B$ and pseudo-magnetic field $B_{S}$. The indexes $(n,m,s,\xi)$
enumerate the eigenstates of the Hamiltonian defined by Eq.(\ref{eq19}), with Landau level $n$, at valley $\xi$ and electronic spin component $s$, for $\lambda=+$. This density matrix operator
is stationary, since in the absence of an external perturbation \cite{vonNeumann} $i\hbar\partial_{t}\hat{\rho} = [\hat{H},\hat{\rho}] = 0$. Here, the coefficient $0 \le p_{n,m,s,\xi}(B)\le 1$
represents the probability for the system, within the statistical ensemble, to be in the particular state
$|\psi_{n,m,s}^{(\xi)}(B)\rangle$. Therefore, the $\{p_{n,m,s,\xi}(B)\}$ satisfy the normalization condition
\begin{eqnarray}
{\rm{Tr}}\hat{\rho} = \sum_{n,m,s,\xi}p_{n,m,s,\xi}(B) = 1.
\label{eq20}
\end{eqnarray}

In the context of Quantum Statistical Mechanics, entropy is defined according
to von Neumann \cite{vonNeumann,Tolman} as $S = -k_{B}{\rm{Tr}}\hat{\rho}\ln\hat{\rho}$.
Since in the energy eigenbasis the equilibrium density matrix operator is diagonal, the entropy reduces to the explicit expression
\begin{eqnarray}
S(B) = -k_{B}\sum_{n,m,s,\xi}p_{n,m,s,\xi}(B)\ln\left(p_{n,m,s,\xi}(B)\right).
\label{eq21}
\end{eqnarray}
In our notation, we emphasize the explicit dependence of the energy eigenstates $\{|\psi_{n,m,s}^{(\xi)}(B)\rangle\}$, as well as the probability coefficients $\{p_{n,m,s,\xi}(B)\}$,
on
the intensity of the external magnetic field $B$.

The ensemble-average energy $E=\langle \hat{H} \rangle$ of the quantum single-particle system is
\begin{eqnarray}
E &=& {\rm{Tr}}(\hat{\rho}\hat{H}) = \sum_{n,m,s,\xi}p_{n,m,s,\xi}(B) E_{n,s}^{\xi}(B)\nonumber\\
&=& \sum_{n,s,\xi}p_{n,s,\xi}(B) E_{n,s}^{\xi}(B),
\label{eq22}
\end{eqnarray}
where we introduced the coefficients
\begin{eqnarray}
p_{n,s,\xi} \equiv \sum_{m} p_{n,m,s,\xi}
\label{eq23}
\end{eqnarray}
in order to take notational advantage of the degeneracy in the spectrum with respect to the quantum number $m$.

The statistical ensemble just described can be submitted to an arbitrary quasi-static process, either by modulating the magnetic field intensity or by exchanging energy with a reservoir. Along such process,
the ensemble-average energy will change accordingly \cite{Munoz_Pena_012,Munoz_Pena_014},
\begin{eqnarray}
dE &=& \sum_{n,s,\xi}p_{n,s,\xi}(B) dE_{n,s}^{\xi}(B) + \sum_{n,s,\xi} dp_{n,s,\xi}(B) E_{n,s}^{\xi}(B)\nonumber\\
&=& \left(\delta E\right)_{\{p_{n,s,\xi}(B)\}={\rm{cnt}}} + \left(\delta E\right)_{\{E_{n,s}^{\xi}(B)\}={\rm{cnt}}}.
\label{eq24}
\end{eqnarray}
The first term in Eq.(\ref{eq24}) represents the energy change due to an iso-entropic process, whereas the second term represents the change due to a process where the energy spectrum remains rigid. These two terms are in correspondence with the macroscopic notions of work and heat, respectively. Therefore, Eq.(\ref{eq24}) represents a microscopic version of the first law of thermodynamics for the statistical ensemble of single-particle systems \cite{Munoz_Pena_012,Munoz_Pena_014},
$dE = \delta W_{\gamma\rightarrow\delta} + \delta Q_{\gamma\rightarrow\delta}$. Here, according to Eq.(\ref{eq24}) the 
work along the process connecting states with magnetic fields $B_{\gamma} \rightarrow B_{\delta}$ is
\begin{eqnarray}
W_{\gamma\rightarrow\delta} &=&\int_{B_{\gamma}}^{B_{\delta}}dB \left( \frac{\partial E}{\partial B}\right)_{\left\{p_{n,s,\xi}(B) \right\}={\rm{cnt}}}\nonumber\\
&=& \int_{B_{\gamma}}^{B_{\delta}}dB \left( \frac{\partial E}{\partial B}\right)_{S}.
\label{eq25}
\end{eqnarray}
On the other hand, the heat exchanged by the system with the environment while modifying its
temperature from $T_{\gamma}\rightarrow T_{\delta}$ will be
\begin{eqnarray}
Q_{\gamma\rightarrow\delta} &=& \int_{T_{\gamma}}^{T_{\delta}}dT\left(\frac{\partial E}{\partial T}  \right)_{B}.
\label{eq26}
\end{eqnarray}

For the statistical ensemble just defined, let us consider a cycle by devising a sequence of quasi-static trajectories in Hilbert's space, as depicted in Fig.\ref{fig2}. Initially, the single-particle system, while submitted to an external magnetic field of intensity $B_{1}$, is brought into thermal equilibrium with a macroscopic thermostat at temperature $T_{1} \equiv T_{C}$. In equilibrium, von Neumann's entropy achieves a maximum, given by the Boltzmann distribution ($\beta=(k_{B} T)^{-1}$)
\begin{eqnarray}
p_{n,s,\xi}(B) = \left[ Z(B,\beta)\right]^{-1} N_{\phi}^{\xi} e^{-\beta E_{n,s}^{\xi}(B)}, 
\label{eq27}
\end{eqnarray}
where the normalization factor is defined by the partition function (see Appendix B for details)
\begin{eqnarray}
Z(B,\beta) = \sum_{n,s,\xi} N_{\phi}^{\xi}e^{-\beta E_{n,s}^{(\xi)}(B)}. 
\label{eq28}
\end{eqnarray}
The ensemble-average energy for the statistical distribution in Eq.(\ref{eq27}) is given by the expression
(see Eq.(\ref{eqB6}) in Appendix B)
\begin{eqnarray}
E(\beta,B) &=& -\left(\frac{\partial \ln Z}{\partial \beta}\right)_{B}.
\label{eq29}
\end{eqnarray}

It also follows from the definition Eq.(\ref{eq21}), along with Eq.(\ref{eq27}) and the normalization condition Eq.(\ref{eq20}) that the entropy can be expressed in terms of the partition function (see Eq.(\ref{eqB7}) in Appendix B)
\begin{eqnarray}
S(\beta,B)/k_{B} &=& \beta E(\beta,B) + \ln Z(B,\beta) \nonumber\\
&=& -\beta \left(\frac{\partial \ln Z}{\partial \beta}\right)_{B} + \ln Z(B,\beta).
\label{eq30}
\end{eqnarray}

The system performs work along the iso-entropic trajectory $1\rightarrow 2$, according to Eq.(\ref{eq25}),
\begin{eqnarray}
W_{1\rightarrow 2} = \int_{B_{1}}^{B_{2}}dB\left( \frac{\partial E}{\partial B}\right)_{S}  = E(T_{2},B_{2}) - E(T_{C},B_{1})
\label{eq31}
\end{eqnarray}
and along the iso-entropic trajectory $3\rightarrow 4$,
\begin{eqnarray}
W_{3\rightarrow 4} = \int_{B_{2}}^{B_{1}}dB\left( \frac{\partial E}{\partial B}\right)_{S}  = E(T_{4},B_{1}) - E(T_{H},B_{2}).\nonumber\\
\label{eq32}
\end{eqnarray}
A physical interpretation of the work performed by the engine is obtained by considering the statistical mechanical definition of the ensemble-average magnetization, that is $M = -\left(\partial E/\partial B \right)_{S}$. Hence, the work terms defined in Eq.(\ref{eq31}) and Eq.(\ref{eq32}) can also be interpreted as $W = -\int M dB$. 

Along the constant magnetic field trajectories $2\rightarrow 3$ and $4\rightarrow 1$, the system exchanges heat with the reservoirs. 
The heat absorbed by the system from the "hot" reservoir at $T_{3} = T_{H}$ is
\begin{eqnarray}
Q_{H} &=& \int_{T_{2}}^{T_{H}}dT\left(\frac{\partial E}{\partial T} \right)_{B_{2}}  \nonumber\\
&=& E(T_{H},B_{2}) - E(T_{2},B_{2}).
\label{eq33}
\end{eqnarray}

Similarly, the heat released by the system to the "cold" reservoir at $T_{C}$ is
\begin{eqnarray}
Q_{C} &=& \int_{T_{4}}^{T_{C}}dT\left(\frac{\partial E}{\partial T} \right)_{B_{1}}  \nonumber\\
&=& E(T_{C},B_{1}) - E(T_{4},B_{1}).
\label{eq34}
\end{eqnarray}

\begin{figure}[tbp]
\centering
\epsfig{file=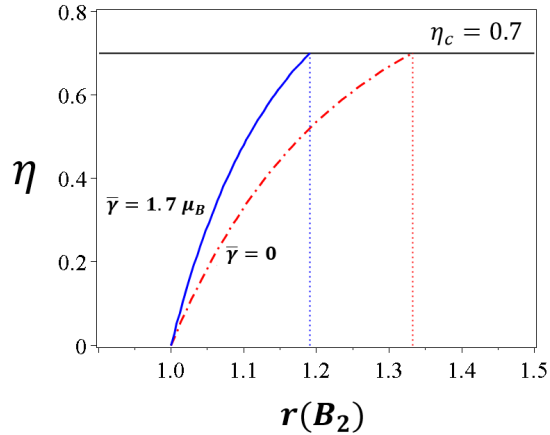,width=0.95\columnwidth,clip=}
\caption{(Color online) The efficiency of the cycle, as a function of the compression ratio $r(B_{2})$, for the case $\bar{\gamma} = 0$ (red, dash-dot line) compared with the case $\bar{\gamma} = 1.7 \mu_{B}$ (blue, solid line). Here $B_{1}=4\,{\rm{T}}$, $B_{S} = 20\,{\rm{T}}$, and the temperatures at the reservoirs $T_{H} = 100\,{\rm{K}}$, $T_{C} = 30\,{\rm{K}}$, respectively.}
\label{fig3}
\end{figure}

The efficiency of the engine is then given by the expression
\begin{eqnarray}
\eta = \left|\frac{W_{1\rightarrow 2} + W_{3\rightarrow 4}}{Q_{H}}   \right | = 1 - \left|\frac{Q_{C}}{Q_{H}}\right|.
\label{eq35}
\end{eqnarray}
The intermediate temperatures $T_{2}$ and $T_{4}$ must be determined numerically from the condition that connects the initial and final states along each iso-entropic trajectory (see Fig.\ref{fig2})
\begin{eqnarray}
S(B_{1},T_{C}) &=& S(B_{2},T_{2}),\nonumber\\
S(B_{2},T_{H}) &=& S(B_{1},T_{4}).
\label{eq36}
\end{eqnarray}

For given values of the initial magnetic field $B_{1}$, the strain pseudo-magnetic field $B_{S}$, and the reservoir temperatures $T_{C}$ and $T_{H}$, the efficiency is a function
of the magnetic field $B_{2}$. We choose to parametrize this dependency by defining the ratio 
\begin{eqnarray}
r(B_{2}) = l_{B_{1}}/l_{B_{2}},
\label{eq37}
\end{eqnarray}
where $l_{B} = {\rm{min}}\{l_{+},l_{-} \}$
is a characteristic confinement length for the semi-classical cyclotronic orbit, defined as the minimum Landau radius
among the two inequivalent valleys. In Fig.~\ref{fig3} we plot the result of
our numerical calculation of the efficiency, as a function of the magnetic field
expressed in terms of the ratio $r(B_{2})$. In this particular example, we have chosen $B_{1} = 4\,{\rm{T}}$, $B_{S}=20\,{\rm{T}}$ and the temperatures $T_{C} = 30\,{\rm{K}}$ and $T_{H}=100\,{\rm{K}}$ at the cold and hot reservoirs, respectively. In Fig.~\ref{fig3}, we also compare the effect of
the pseudo-Zeeman term, by calculating the efficiency when setting $\bar{\gamma} = 1.7 \mu_{B}$ and $\bar{\gamma}=0$, respectively. It is evident from comparison of both curves that the pseudo-Zeeman effect produces a relative enhancement of the efficiency as compared to the case when this term is absent.

We find that the numerical solution for the efficiency as a function of the compression ratio can be accurately represented by the parametric form
\begin{eqnarray}
\eta = 1 - [r(B_{2})]^{-\alpha}.
\label{eq38}
\end{eqnarray}
Here, the exponent $\alpha$ depends on the temperatures $T_{C}$ and $T_{H}$, as well as on the strain field $B_{S}$. In particular, for the choice of parameters represented in Fig.\ref{fig3}, we find that $\alpha=6.88$ for $\bar{\gamma}=1.7\mu_{B}$, whereas $\alpha = 4.2$ for $\bar{\gamma}=0$.
Remarkably, this parametric expression is analogous to the well known formula for the
efficiency of the Otto cycle that works with a classical ideal gas, with $r$ instead of the volumetric ratio that applies to the classical case. An even closer analogy between both cases can be put forward by
recognizing that $r(B_{2}) = l_{B_{1}}/l_{B_{2}} > 1$ can be literally interpreted as a "compression ratio" between the effective Landau radii, which in practice defines a characteristic confinement length for the semi-classical cyclotronic orbit associated to each Landau level.

Numerical solutions for the efficiency only exist up to a maximum compression ratio, which in the example displayed in Fig.~\ref{fig3} is $r^{max} =  1.19$ for $\bar{\gamma} = 1.7\mu_{B}$, whereas
$r^{max}=1.33$ for $\bar{\gamma}=0$. At this point, the efficiency attains its maximum value, that exactly matches the Carnot efficiency for the same temperatures in the thermostats, i.e. $\eta(r^{max})=\eta_{C} = 1 - T_{C}/T_{H} = 0.7$. More generally, using the parametric form Eq.(\ref{eq38}), one concludes that
\begin{eqnarray}
r < \left(\frac{T_{H}}{T_{C}}\right)^{1/\alpha} \equiv r^{max}.
\label{eq39}
\end{eqnarray}

\section{Concluding Remarks}
We believe that for a practical implementation of the cycle proposed in this work, it should be more convenient to apply a combination of a real magnetic field and strain, than with strain solely. Experimentally, it is easier to impose a dynamic modulation of the applied magnetic field intensity, rather than to modulate strain over a nanoscale sample. In our proposed scheme, only a static strain pattern is imposed. On the other hand, using only a real magnetic field in the absence of mechanical strain, presents the disadvantage of requiring very high magnetic field intensities. This problem is avoided by the strain induced pseudo-magnetic field, which experimentally can achieve up to $B_{S} \sim 300\,{\rm{T}}$ \cite{Levy_2010}, and thus provides a high baseline for the total effective field $|B_{\pm}| = |B \pm B_{S}|$ that allows for the emergence of relativistic Landau levels, which as we propose can be modulated with a real magnetic field $B$ of moderate intensity.

\section*{Acknowledgements}

E.M. acknowledges financial support from Fondecyt Grant 1141146. F.J.P.
acknowledges financial support from a Conicyt fellowship.

\appendix

\section{}
In this appendix, we provide details of the solution for the single-particle spectrum of the Hamiltonian Eq.(\ref{eq17}), following the operator method presented in \cite{Goerbig_2011}. Let us first define the canonical momentum operators at each valley $K_{\xi}$,
\begin{eqnarray}
\mathbf{\Pi}^{\xi} = -i\hbar\nabla + e\left(\mathbf{A} + \xi\mathbf{A}_{S} \right). 
\label{eqA1}
\end{eqnarray}
The single particle Hamiltonian at each valley $K_{\xi}$, for $\xi=\pm$ is therefore expressed by
\begin{eqnarray}
\hat{H}^{\xi} = \xi v_{F}\hat{\boldsymbol{\sigma}}\cdot\mathbf{\Pi}^{\xi}  - \bar{\gamma}\xi B_{S}\hat{\sigma}_{3}
-s \gamma B \mathbf{1}.
\label{eqA2}
\end{eqnarray}
The eigenvalue problem can then be formulated in the  form
\begin{eqnarray}
\left( \xi v_{F} \hat{\boldsymbol{\sigma}}\cdot\mathbf{\Pi}^{\xi}  - \bar{\gamma}\xi B_{S}\hat{\sigma}_{3} \right)
|\psi_{s}^{(\xi)}\rangle = \epsilon_{s}^{\xi}|\psi_{s}^{(\xi)}\rangle,
\label{eqA3}
\end{eqnarray}
where we have defined $\epsilon_{s}^{\xi} = E_{s}^{\xi} + s\gamma B$.
Let us now define the Landau radius associated to the total effective magnetic field $B_{\xi} = B + \xi B_{S}$ at each valley,
\begin{eqnarray}
l_{\xi} = \sqrt{\frac{\hbar}{e\left| B_{\xi}\right|}}.
\label{eqA4}
\end{eqnarray}
It is straightforward to verify that the components of the canonical momentum operator satisfy the
commutation relation
\begin{eqnarray}
\left[ \Pi_{x}^{\xi}, \Pi_{y}^{\xi}  \right] = -i\frac{\hbar^{2}}{l_{\xi}^{2}} \sgn (B_{\xi}),
\label{eqA5}
\end{eqnarray}
which allows us to introduce the creation and annihilation operators 
\begin{eqnarray}
\hat{a} &=& \frac{l_{\xi}}{\hbar \sqrt{2}} \left( \Pi_{x}^{\xi} - i \sgn (B_{\xi}) \Pi_{y}^{\xi} \right),\nonumber\\
\hat{a}^{\dagger} &=& \frac{l_{\xi}}{\hbar \sqrt{2}} \left( \Pi_{x}^{\xi} + i \sgn (B_{\xi}) \Pi_{y}^{\xi} \right),
\label{eqA6}
\end{eqnarray}
satisfying the canonical commutation relation
\begin{eqnarray}
\left[ \hat{a},\hat{a}^{\dagger}\right] = 1
\label{eqA7}
\end{eqnarray}
as a consequence of Eq.(\ref{eqA6}).
In explicit matrix notation, the eigenvalue problem then reduces to (for $\sgn (B_{\xi}) >0$)
\begin{eqnarray}
\xi\hbar\Omega_{\xi}\left[\begin{array}{cc}-\Delta_{\xi} & \hat{a}\\\hat{a}^{\dagger} & \Delta_{\xi}\end{array} \right]
\left(\begin{array}{c}|u_{n}\rangle \\ |v_{n}\rangle\end{array} \right) = \epsilon_{n,s}^{\xi}\left(\begin{array}{c} |u_{n}\rangle \\ |v_{n}\rangle\end{array} \right),
\label{eqA8}
\end{eqnarray}
with $\Delta_{\xi} = \bar{\gamma}B_{S}\hbar^{-1}\Omega_{\xi}^{-1}$.
The matrix structure for $\sgn (B_{\xi})< 0$ is obtained under the similarity transformation $\hat{H}^{\xi} \rightarrow \hat{\sigma}_{1}\hat{H}^{\xi}\hat{\sigma_{1}}$, and hence it shares exactly the same spectrum. After Eq.(\ref{eqA8}), we have the eigensystem of equations
\begin{eqnarray}
\xi\hbar\Omega_{\xi}\hat{a}|v_{n}\rangle = \left(\epsilon_{n,s}^{\xi} + \xi\hbar\Omega_{\xi}\Delta_{\xi} \right)|u_{n}\rangle,
\label{eqA9}\\
\xi\hbar\Omega_{\xi}\hat{a}^{\dagger}|u_{n}\rangle = \left(\epsilon_{n,s}^{\xi} - \xi\hbar\Omega_{\xi}\Delta_{\xi} \right)|v_{n}\rangle.
\label{eqA10}
\end{eqnarray}
For the ladder operators defined by Eqs.(\ref{eqA6}) and (\ref{eqA7}), the combination $\hat{a}^{\dagger}\hat{a} = \hat{n}$ is the number operator. Combining Eq.(\ref{eqA9}) and Eq.(\ref{eqA10}), we have
\begin{eqnarray}
\hat{a}^{\dagger}\hat{a}|v_{n}\rangle = \frac{\left(\epsilon_{n,s}^{\xi}\right)^{2}- \left(\hbar\Omega_{\xi}\Delta_{\xi}\right)^{2}}{\left( \hbar\Omega_{\xi}\right)^{2}}|v_{n}\rangle.
\label{eqA11}
\end{eqnarray}
Clearly then, $|v_{n}\rangle = |n\rangle$ is an eigenstate of the number operator, $\hat{a}^{\dagger}\hat{a}|v_{n}\rangle = \hat{a}^{\dagger}\hat{a}|n\rangle = n|v_{n}\rangle$. Therefore, from
Eq.(\ref{eqA11}) we obtain
\begin{eqnarray}
\epsilon_{\lambda,n,s}^{\xi} = \lambda\hbar\Omega_{\xi}\sqrt{n + \Delta_{\xi}^{2}},\,\,\,\,n>0,
\label{eqA12}
\end{eqnarray}
where the sign $\lambda=\pm$ arising from the two roots of the quadratic equation determines the valence and conduction bands, respectively. The first component $|u_{n}\rangle$ of the spinor, for $n>0$, is determined from Eq.(\ref{eqA9}), using $\hat{a}|v_{n}\rangle = \sqrt{n}|n-1\rangle$, as follows
\begin{eqnarray}
|u_{n}\rangle &=& \frac{\xi\hbar\Omega_{\xi}}{\epsilon_{n,s}^{\xi}+\xi\hbar\Omega_{\xi}\Delta_{\xi}}\hat{a}|v_{n}\rangle\nonumber\\
&=& \frac{\sqrt{n}}{\Delta_{\xi}+ \lambda\xi\sqrt{n+\Delta_{\xi}^{2}}}|n-1\rangle
\label{eqA12b}
\end{eqnarray}

The corresponding (normalized) spinor eigenvectors
are thus given by
\begin{eqnarray}
|\psi_{\lambda,n,s}^{(\xi)}\rangle = \frac{1}{\sqrt{2}}\left( \begin{array}{c} \alpha_{n}|n-1\rangle \\ \beta_{n} |n\rangle\end{array}\right),\,\,\,n>0,
\label{eqA13}
\end{eqnarray}
where we have defined the coefficients
\begin{eqnarray}
\alpha_{n} &=& \frac{\sqrt{n}}{\left(n + \Delta_{\xi}^{2} + \lambda\xi\Delta_{\xi}\sqrt{n + \Delta_{\xi}^{2}}\right)^{1/2}},\nonumber\\
\beta_{n} &=& \frac{\Delta_{\xi}+ \lambda\xi\sqrt{n + \Delta_{\xi}^{2}}}{\left(n + \Delta_{\xi}^{2} + \lambda\xi\Delta_{\xi}\sqrt{n + \Delta_{\xi}^{2}}\right)^{1/2}}.
\label{eqA13b}
\end{eqnarray}
The case $n=0$ must be analyzed separately. From Eq.(\ref{eqA9}), we conclude that 
$|u_{n}\rangle \propto \hat{a}|v_{n}\rangle \propto |n-1\rangle$. In particular, since $|v_{n=0}\rangle = |0\rangle$, then it follows that $|u_{n=0}\rangle = 0$. Applying this condition to Eq.(\ref{eqA10})
for $n=0$, one immediately concludes
\begin{eqnarray}
\epsilon_{0,s}^{\xi} = \xi\hbar\Omega_{\xi}\Delta_{\xi}.
\label{eqA14}
\end{eqnarray}
The corresponding spinor eigenvector for the Landau level $n=0$ is thus given by
\begin{eqnarray}
|\psi_{n=0,s}^{(\xi)}\rangle = \left( \begin{array}{c} 0 \\ |0\rangle\end{array}\right),\,\,\,n=0.
\label{eqA15}
\end{eqnarray}
The spectrum is then obtained by the shift $E_{\lambda,n,s}^{\xi} = \epsilon_{\lambda,n,s}^{\xi} - s\gamma B$,
\begin{eqnarray}
E_{\lambda,n,s}^{\xi} = \left\{ \begin{array}{cc}\lambda\hbar\Omega_{\xi}\sqrt{n + \Delta_{\xi}^{2}} - s\gamma B, & n > 0,\\
\xi \hbar\Omega_{\xi} \Delta_{\xi} - s\gamma B, & n=0.\end{array}\right.
\label{eqA16}
\end{eqnarray}

So far, we have only considered the quantization of the orbital degrees of freedom. However, an intrinsic (very large) degeneracy in the eigenvalues described by Eq.(\ref{eqA16}) emerges when we take into
account the quantization of the {\it{guiding-center}} of the orbit, i.e. the center of the classical cyclotron
orbit. For that purpose, one decomposes the position operator $\mathbf{r} = (x,y)= \mathbf{R} + \boldsymbol{\rho}$ into the guiding-center coordinate $\mathbf{R} = (X,Y)$ and the cyclotron orbital variable $\boldsymbol{\rho} = (\rho_{x},\rho_{y})$. The later is perpendicular to the electron's velocity, and hence it is expressed in terms of the
canonical momentum components \cite{Goerbig_2011}
\begin{eqnarray}
\rho_{x} = \frac{\Pi_{y}^{\xi}}{eB_{\xi}},\,\,\,\,\rho_{y} = -\frac{\Pi_{x}^{\xi}}{eB_{\xi}}.
\label{eqA17}
\end{eqnarray}
In consequence, one has the commutation relations that follow after Eq.(\ref{eqA5})
\begin{eqnarray}
\left[  \rho_{x}, \rho_{y} \right] = \frac{\left[ \Pi_{x}^{\xi},\Pi_{y}^{\xi}  \right]}{\left(e B_{\xi}\right)^{2}} = -i l_{\xi}^{2}
\sgn(B_{\xi}).
\label{eqA18}
\end{eqnarray}
The commutation between the components of the position operator $[x,y] = 0$ implies the commutation
relation
\begin{eqnarray}
\left[ X,Y\right] = -\left[\rho_{x}, \rho_{y}\right] = i l_{\xi}^{2}\sgn(B_{\xi}).
\label{eqA19}
\end{eqnarray}
Therefore, we are entitled to define a second pair of creation and annihilation operators
\begin{eqnarray}
\hat{b} &=& \frac{1}{\sqrt{2}l_{\xi}}\left( X + i \sgn(B_{\xi})Y  \right),\nonumber\\
\hat{b}^{\dagger} &=& \frac{1}{\sqrt{2}l_{\xi}}\left( X - i\sgn(B_{\xi}) Y  \right)
\label{eqA20}
\end{eqnarray}
that satisfy the canonical commutation relation $[\hat{b},\hat{b}^{\dagger} ] = 1$, thus defining
a pair of ladder operators, characterized by the eigenvalue equation
\begin{eqnarray}
\hat{b}^{\dagger}\hat{b}|m\rangle = m|m\rangle.
\label{eqA21}
\end{eqnarray}
Therefore, the full solution of the eigenvalue problem stated in Eq.(\ref{eqA3}) is obtained by the tensor product of the spinor eigenstates defined in Eqs.(\ref{eqA13}), (\ref{eqA15}) and the 
eigenstates $|m\rangle$ associated to the guiding-center coordinates
\begin{eqnarray}
|\psi_{\lambda,n,m,s}^{(\xi)}\rangle &=& |\psi_{\lambda,n,s}^{(\xi)}\rangle \otimes |m\rangle = \frac{1}{\sqrt{2}}\left( \begin{array}{c} \alpha_{n} |n-1,m\rangle \\ \beta_{n} |n,m\rangle\end{array}  \right),\,\,n>0\nonumber\\\nonumber\\
|\psi_{n=0,m,s}^{(\xi)}\rangle &=& |\psi_{n=0,s}^{(\xi)}\rangle \otimes |m\rangle = \left( \begin{array}{c} 0 \\  |0,m\rangle\end{array}  \right),\,\,n=0.\nonumber\\
\label{eqA22}
\end{eqnarray}
Here, the coefficients $\alpha_{n}$ and $\beta_{n}$ are as defined in Eq.(\ref{eqA13b}).
The uncertainty relation between the guiding-center coordinates that follows from the commutation relation Eq.(\ref{eqA19}) is \cite{Goerbig_2011}
\begin{eqnarray}
\Delta X \Delta Y = 2\pi l_{\xi}^{2}.
\label{eqA23}
\end{eqnarray}
Just as in the case of phase-space coordinates, this uncertainty relation defines the minimal area of an elementary surface cell associated to a given eigenstate. Therefore, the degeneracy $N_{n,s}^{\xi}$ of each Landau level, corresponding to the maximum value of $m$ allowed, is given by the number of elementary cells contained in the total
surface $\mathcal{A}$ of the sample
\begin{eqnarray}
N_{n,s}^{\xi} = N_{\phi}^{\xi} = \frac{\mathcal{A}}{\Delta X \Delta Y} = \frac{\mathcal{A}}{2\pi l_{\xi}^{2}} = \frac{\Phi_{\xi}}{\phi_{0}}.
\label{eqA24}
\end{eqnarray}
Here, $\Phi_{\xi} = B_{\xi}\mathcal{A}$ is the flux of the "total" effective magnetic field piercing the
area of the sample, whereas $\phi_{0} = h/e$ is the universal magnetic flux quantum. 
Notice that in general, the degeneracy factors for a given external magnetic field $B$ and strain pseudo-magnetic field $B_{S}$ will be different at each valley $K_{\xi}$.
\section{}
The single-particle spectrum corresponds to the relativistic Landau levels for the effective magnetic
field $B_{\xi} = B + \xi B_{S}$, involving the strain $B_{S}$ at each valley and the real magnetic field $B$ contributions
\begin{eqnarray}
E_{n,s}^{\xi} = \left\{ \begin{array}{cc}\hbar\Omega_{\xi}\sqrt{n + \Delta_{\xi}^{2}} - s\gamma B, & n > 0,\\
\xi \hbar\Omega_{\xi} \Delta_{\xi} - s\gamma B, & n=0,\end{array}\right.
\label{eqB1}
\end{eqnarray}
for $\Delta_{\xi} = \bar{\gamma}B_{s}\hbar^{-1}\Omega_{\xi}^{-1}$. Here, we consider only particle-like solutions ($\lambda > 0$), assuming that the graphene flake  has been
charged with a single electron in the conduction band.
The partition function that describes the statistical ensemble of single-particle systems is given
by
\begin{eqnarray}
Z &=& \sum_{\xi=\pm}\sum_{s=\pm}\sum_{n=0}^{\infty}\sum_{m=0}^{N_{n,s}^{\xi}-1}  e^{-\beta E_{n,s}^{\xi}}\nonumber\\
&=& \sum_{\xi=\pm}\sum_{s=\pm}\sum_{n=0}^{\infty} N_{n,s}^{\xi} e^{-\beta E_{n,s}^{\xi}}.
\label{eqB2}
\end{eqnarray}
 The degeneracy 
of each relativistic Landau level $n \ge 0$ with respect to the center of mass quantum number $m\ge 0$ corresponds to the number of effective magnetic flux quanta
piercing the area $\mathcal{A}$ of the sample, $N_{n,s}^{\xi} = N_{\phi}^{\xi} = \Phi_{B_{\xi}}/\phi_{0}$.
Under these considerations, the partition
function factors into the form
\begin{eqnarray}
Z &=& 2 \cosh(\beta\gamma B) \left( N_{\phi}^{+} Z_{+} + N_{\phi}^{-} Z_{-} \right). 
\label{eqB3}
\end{eqnarray}
Here, we have defined
\begin{eqnarray}
Z_{\xi} = e^{-\xi \beta \hbar \Omega_{\xi} \Delta_{\xi}} + \sum_{n=1}^{\infty} e^{-\beta \hbar\Omega_{\xi}\sqrt{n + \Delta_{\xi}^{2}}}.
\label{eqB4}
\end{eqnarray}
For the low temperatures of interest, and
at relatively large magnetic fields, one finds that at both valleys $\beta \hbar\Omega_{\xi} \gg 1$. For example, consider $T \sim 100$ K
and $B_{\xi} \sim 10$ T, which are of the order of magnitude of the numerical examples discussed in the main text. For this case, one readily finds $\beta \hbar\Omega_{\xi} \sim 13.3$. Given this fact, the exponential terms
which add up to the infinite series are very small, and hence the sum can be well approximated within less than $2\%$ of relative error by retaining only
the first term,
\begin{eqnarray}
Z_{\xi} \sim e^{-\beta \bar{\gamma}\xi B_{S}} + e^{-\beta \hbar\Omega_{\xi}\sqrt{1 + \Delta_{\xi}^{2}}}.
\label{eqB5}
\end{eqnarray}
Using this expression, explicit analytical formulas are obtained for the ensemble-average energy $E$
\begin{widetext}
\begin{eqnarray}
E(\beta,B) = -\left(\frac{\ln Z}{\partial\beta}\right)_{B}
= -\gamma B\tanh(\beta\gamma B) + \frac{\sum_{\xi=\pm}N_{\phi}^{\xi}\left(\xi\bar{\gamma} B_{S}
e^{-\xi\beta\bar{\gamma}B_{S}} + \hbar\Omega_{\xi}\sqrt{1+\Delta_{\xi}^{2}} 
e^{-\beta\hbar\Omega_{\xi}\sqrt{1+ \Delta_{\xi}^{2}}} \right)}{\sum_{\xi=\pm} N_{\phi}^{\xi}\left(  e^{-\xi\beta\bar{\gamma}B_{S}}
+  e^{-\beta\hbar\Omega_{\xi}\sqrt{1+ \Delta_{\xi}^{2}}}  \right)},
\label{eqB6}
\end{eqnarray}
\end{widetext}
as well as for the entropy $S / k_{B} = \ln Z -\beta\frac{\partial\ln Z}{\partial\beta}$
\begin{eqnarray}
&&\frac{S(\beta,B)}{k_{B}} = \beta E(\beta,B) + \ln\left[2\cosh(\beta\gamma B)\right]
\label{eqB7}
\\
&+& \ln\left[\sum_{\xi=\pm} N_{\phi}^{\xi}\left(  e^{-\xi\beta\bar{\gamma}B_{S}}
+  e^{-\beta\hbar\Omega_{\xi}\sqrt{1+ \Delta_{\xi}^{2}}}  \right) \right].\nonumber
\end{eqnarray}

\bibliographystyle{apsrev}

\end{document}